\documentclass[a4paper,12pt]{JHEP3}
\usepackage{amsmath}
\preprint{SISSA 19/2006/EP}
\numberwithin{equation}{section}

\allowdisplaybreaks
\title{\Large The  $\alpha'$ stretched horizon in the Heterotic string}

\author{{Ghasem Exirifard}\\
SISSA/ISAS, Via Beirut 2-4, I-34013 Trieste, Italy\\INFN, Sezione di Trieste, Italy\\
 Email: \email{exir@sissa.it}
}

\abstract{The linear $\alpha'$ corrections and the field redefinition ambiguities are studied for half-BPS singular backgrounds representing a wrapped fundamental string. It is showed that there exist schemes in which the inclusion of all the linear $\alpha'$ corrections converts these singular solutions to black holes with a regular horizon for which the modified  Hawking-Bekenstein entropy is in agreement with the statistical entropy.}


\begin{document}
\bibliographystyle{JHEP}

\section{Introduction}
The massless field of helicity two in the spectrum of string theory is identified as the gravitational field since its low energy effective action around flat space-time coincides with the Einstein-Hilbert action. This identification sets the subleading string corrections as the quantum corrections to gravity and allows one to ask if and how quantum corrections preserve or change the properties of  classical backgrounds. In particular one may ask if the subleading string corrections induce a regular horizon on the singular classical geometries which have an entropy associated to them. 

Amongst these singular classical geometries are the half BPS null singular ones which represent a wrapped fundamental string with general momentum and winding numbers \cite{peet}. These null singular geometries have a statistical entropy associated to them since string states with given momentum and winding numbers are degenerate \cite{atish}. It is conjectured that quantum effects convert these singular geometries to black holes with a regular horizon. 

It is known that the the leading world-sheet corrections of the Heterotic string includes the square of the Riemann tensor.  Ref \cite{atish2}, motivated by \cite{vafaatractor}, observed that the inclusion of the square of the Riemann tensor  and its supersymmetric partners in $D=4$ \cite{19,20,21,22,23,24,25,26,27,28} induces a local horizon with geometry $AdS_2\times S^2$ on these backgrounds and for which the modified Hawking-Bekenstein entropy \cite{wald,wald2,Myers} is in agreement with the statistical entropy. This observation renewed interest in the subject \cite{int1,int2,int3,int5,Cardoso:2006nt,Dabholkar:2005dt,Sahoo:2006rp,Alishahiha:2006ke,Chandrasekhar:2006zw,Ghodsi:2006cd}. Ref. \cite{int5,sen,sen0} introduced the entropy formalism and concluded that the inclusion of the Gauss-Bonnet action as a part of the linear $\alpha'$ corrections in an arbitrary dimension induces a local horizon with geometry $AdS_2\times S^{D-2}$ for which the modified Hawking-Bekenstein entropy is in agreement with the statistical entropy up to a numerical constant factor.

In this note we present a way to calculate all the linear $\alpha'$ corrections in an arbitrary dimension and we study how these corrections may change these null singular backgrounds to black holes. The note is organised in the following way;

In the second section we review the  classical solutions representing a wrapped fundamental string on a two cycle. We realise them as  ten dimensional backgrounds composed of the metric, the NS two form and the dilaton first compacted on a torus of appropriate dimensionality to $D+1$ dimensional space-time and then through KK compactification on a circle to a $D$ dimensional space-time.

In the third section we review how the $\alpha'$ corrections can be computed. We present the linear $\alpha'$ corrections in the Heterotic theory to backgrounds of metric, NS two form and dilaton obtained from scattering amplitude considerations \cite{stringamplitude1,stringamplitude2}. We study the field redefinition ambiguities. We require that the generalisation of the Einstein tensor is covariantly divergence free. This requirement fixes the curvature squared terms to the Gauss-Bonnet Lagrangian keeping some of the field redefinition ambiguity parameters untouched.

In the fourth section we discuss how the singularity could be modified by the inclusion of the $\alpha'$ corrections. We employ the compactification process of the first section to account for all the linear $\alpha'$ corrections in lower dimensions using the corrections in ten dimensions. We compute the local horizon configuration parameters for all field redefinitions compatible with ten dimensional diffeomorphism group. Note that the modified Hawking-Bekenstein entropy is the same for actions related to each other by field redefinition provided that  the $\alpha'$ terms are studied as perturbations around a classical solution \cite{Jacobson:1993vj}. However since the local horizon is the exact solution of the truncated equations then the modified Hawking-Bekenstein entropy depends on the field redefinition ambiguity parameters. We show that there exist schemes in which the inclusion of all the linear $\alpha'$ corrections in an arbitrary dimension gives rise to a local horizon with geometry $AdS_2\times S^{D-2}$ for which the modified Hawking-Bekenstein entropy is in agreement with the statistical entropy and outside which the higher order $\alpha'$ corrections are perturbative. We also discuss on the existence of a smooth solution connecting the local horizon to asymptotic infinity.

In the fifth section the conclusions are presented. 

\section{The tree-level singular background}
The low energy effective action of the critical heterotic string theory for the metric ($\boldsymbol{g}$), the NS two-form ($\boldsymbol{B}$) and the dilaton ($\boldsymbol{\phi}$ ) reads
\begin{eqnarray}\label{treelevelaction}
\boldsymbol{S}^{(10)}&=&\frac{1}{32\, \pi}\,\int d^{10}\boldsymbol{x}~\sqrt{-\boldsymbol{g}}\,e^{-2\boldsymbol{\phi}}\,\boldsymbol{L}^{\scriptscriptstyle(10)}\\
\boldsymbol{L}^{\scriptscriptstyle(10)}&=&({\boldsymbol{R}_{\scriptscriptstyle{\text{Ricci}}}} + 4 |\boldsymbol{\nabla} \boldsymbol{\phi}|^2 - \frac{1}{12} \boldsymbol{H}_{ijk}\boldsymbol{H}^{ijk})\,,
\end{eqnarray}
where
\begin{eqnarray}\label{defineH}
\boldsymbol{H}_{ijk}&=& {3} \boldsymbol{B}_{[ij,k]}\,.
\end{eqnarray}
The bold symbols will be used to represent the fields in ten dimensions. Note that we are not using the modified field strength \cite{DBranes}
\begin{equation}\label{convetionalH}
\boldsymbol{H_{\scriptscriptstyle\text{modified}}}=d\boldsymbol{B}-\frac{\alpha'}{4}[\frac{1}{30}\boldsymbol{\omega}_{3Y}(A)-\boldsymbol{\omega}_{3L}(\Omega)]\,,
\end{equation}
 where $\boldsymbol{\omega}_{3Y}(A)$ and $\boldsymbol{\omega}_{3L}(\Omega)$  stand for the Chern-Simons three-forms associated respectively to either the $\text{Spin}(32)/Z_2$ or $E8 \times E8$ connection and to the spin connection. In this work we are considering backgrounds of vanishing gauge connections where  $\boldsymbol{\omega}_{3Y}(A)=0$.  The $\alpha'$ term in (\ref{convetionalH}) represents a part of the linear $\alpha'$ corrections identified during the study of the anomaly cancellation. We use (\ref{defineH}) and in the next section we will add all the linear $\alpha'$ corrections.

We are interested in the extrema of (\ref{treelevelaction}) whose fields configuration follows 
\begin{eqnarray}
{\boldsymbol{ds^2}}&=&\sum_{\scriptscriptstyle{\mu,\nu=1}}^{D} \boldsymbol{g}_{\mu\nu} (x) dx^{\mu} dx^{\nu} + 2 \boldsymbol{g}_{y\mu}(x)\,dy\, dx^{\mu} + \boldsymbol{g}_{yy}(x) dy^2 +\sum_{\scriptscriptstyle{m=D+1}}^{10} dz_{m}^2\,,\\
{\boldsymbol{B}}&=&\boldsymbol{B}_{\mu\nu}(x)dx^{\mu}\wedge dx^{\nu}+\boldsymbol{B}_{y\mu}(x) dx^{y}\wedge dy\,,\\
{\boldsymbol{\phi}}&=&\boldsymbol{\phi}(x)\,,\\
\label{ycordinate}
y&\sim& y + 8 \pi \,, 
\end{eqnarray}
where $y$ is compactified on a circle and $z_i$ are compactified on $T^{9-D}$. These extrema are examples of trivial compactification on a torus of appropriate dimensionality from ``$10$" dimensions to a ``$D+1$" dimensional space-time and then KK compactification on a circle  to a $D$ dimensional space-time. If one represents non-trivial components of the ten dimensional fields by
\begin{eqnarray}\label{compactification}
\begin{array}{rcl}
\boldsymbol{g}_{yy}(x) &=& T^2\,,\\
\boldsymbol{g}_{\mu\nu}(x) &=& g_{\mu\nu}+ 4 T^2 A_{\mu}^{\scriptscriptstyle{(1)}} A_{\nu}^{\scriptscriptstyle{(1)}}\,,\\
2 \boldsymbol{\phi}(x)&=&2 \phi + \ln T\,-\ln V,
\end{array}&~&
\begin{array}{rcl}
\boldsymbol{g}_{y\mu}(x) &=& 2 A^{{(1)}} T^2\,,\\
\boldsymbol{B}_{y\mu}(x) &=& 2 A^{{(2)}}_{\mu}\,,\\
\boldsymbol{B}_{\mu\nu}(x) &=& B_{\mu\nu} + 2 (A_{\mu}^{\scriptscriptstyle{(1)}}A_{\nu}^{\scriptscriptstyle{(2)}}-A_{\nu}^{\scriptscriptstyle{(1)}}A_{\mu}^{\scriptscriptstyle{(2)}})\,,
\end{array}
\end{eqnarray}
where $V$ is the volume of the compact directions. Then the induced action for the new fields - $g, A^{\scriptscriptstyle{(1)}},A^{\scriptscriptstyle{(2)}}, B, T$ and $\phi$- reads
\begin{eqnarray}\label{inducedaction}
{S}&=&\int d^{D}x {L}\\
&=&\,\frac{1}{32\pi}\int d^Dx \sqrt{-g} e^{-2\phi}(R_{\scriptscriptstyle{\text{Ricci}}}+4|\nabla \phi|^2 - \frac{|\nabla T|^2}{T^2}-\frac{|dB|^2}{12} - T^2 |dA^{\scriptscriptstyle{(1)}}|^2 - \frac{|dA^{\scriptscriptstyle{(2)}}|^2}{T^2}),\nonumber
\end{eqnarray}
where $R_{\scriptscriptstyle{\text{Ricci}}}$ is the Ricci scalar of $g_{\mu\nu}$, and an integration by parts is understood
\begin{equation}
\boldsymbol{L}^{\scriptscriptstyle(10)}- {L}\,=\,2\sqrt{-g}~\nabla^{\mu}(e^{-2\phi}\,{\frac{\nabla_{\mu}T}{T}})\,.
\end{equation}
We refer to (\ref{inducedaction}) as the induced action, and to $x_\mu$ and ($z^\mu$,$y$) respectively as the large dimensions and as the compactified space. Due to the form of the induced action it is natural to interpret $A^{\scriptscriptstyle{(1)}}$ and $A^{\scriptscriptstyle{(2)}}$
 as different $U(1)$ gauge connections in the large dimensions.\footnote{Ref.  (\cite{Maharana:1992my}) shows that the fields in large dimensions should be defined by (\ref{compactification})  in order to not   mix the U(1) symmetries.} A family of the extrema of the compactified action is given by 
\begin{eqnarray}\label{background1}
ds^2_{\scriptscriptstyle{\text  string}}&=&-\,e^{4\phi(r)}\,dt^2\,+\,dr^2\,+\,r^2\,d\Omega_{D-2}^2\,,\\\nonumber\\
e^{-4\phi(r)}&=&\frac{{(r^{D-3}+2W)\,(r^{D-3}+2N)}}{r^{2(D-3)}}\,,\qquad \label{background2}
T(r)\,=\,\sqrt{\frac{r^{D-3}+2N}{r^{D-3}+2W}}\,,\\
A^{\scriptscriptstyle{(1)}}_\tau(r)&=&-\,\frac{N}{r^{D-3}+2N}\,,\hspace*{3.6cm} \label{background3}
A^{\scriptscriptstyle{(2)}}_\tau(r)\,=\,-\,\frac{W}{r^{D-3}+2W}\,,
\end{eqnarray}
where $N$ and $W$ are two arbitrary numbers labelling the solution. We only consider the case where $N$ and $W$ are both positive. These backgrounds are constructed in \cite{peet} as singular limits of regular black-holes obtained by applying a solution generating transformation \cite{sol1,sol2} on a higher dimensional Kerr metric. Here we use the notation of \cite{sen2}. Ref. \cite{peet} proved that they break half of the ten dimensional supersymmetries leaving eight unbroken supersymmetry parameters. These backgrounds are null-singular, i.e. the horizon coincides with the singularity. They represent  BPS states of an elementary string carrying $n$ units of momentum and $w$ units of winding charges along $S^1$ of the $y$ coordinate where \cite{sen2}
\begin{eqnarray}
n&=&\frac{(D-3)\Omega_{D-2}}{4\pi}\,N\,,\\
w&=&\frac{(D-3)\Omega_{D-2}}{4\pi}\,{W}\,,
\end{eqnarray}
and the unit of $\alpha'=16$ is used.\footnote{We have chosen a specific value for the radius of the compactification because the $\alpha'$ perturbative corrections to (\ref{background1}) do not depend on the radius of the compactification. The solution which represents KK-compactification on a circle with an arbitrary radius can be generated by rescaling $y$ and using  (\ref{compactification}). This solution is  written in \cite{sen2}.} For general values of $N$ and $W$ a tachyon instability may exist around the singularity, reminiscent of the tachyon instability outside the horizon of Euclidean black holes presented in \cite{tachyon,tachyon2}. We focus on the cases where $N\sim W$ and this instability is not present.

An entropy may be associated to these backgrounds since in general there exists more than one state of the Heterotic string carrying $w$ units of winding and $n$ units of momentum along $S^1$ of the $y$ coordinate. For large $n$ and $w$ the degeneracy of these states grows as $e^{4\pi\sqrt{nw}}$ \cite{atish1}. Thus the entropy, defined by the logarithm of the degeneracy of the states, is given by:
\begin{equation}\label{statisticalentropy}
S_{\text{statistical}}\,=\,4\pi\sqrt{nw}\,,
\end{equation}
when $n$ and $w$ are large. We refer to this entropy as the statistical entropy. A dilemma will arise as soon as the statistical entropy is associated to these tree-level backgrounds since they are singular and do not possess a regular event horizon to which the thermodynamical properties can be connected. This dilemma can be resolved in either of the following ways,
\makeatletter
\renewcommand{\theenumi}{\Roman{enumi}} 
\makeatother
\begin{enumerate}\label{resulotions}
\item Statistical entropy should not be associated to these backgrounds.
\item Thermodynamical properties should be expressed in term of something else, in place of the event horizon, which null-singular geometries possess.
\item The subleading string corrections will induce an event horizon and the horizon cloaks the singularity.
\end{enumerate}
Of the above possibilities, the first seems unnatural since  the statistical entropy is associated to  regular black holes \cite{vafa,vafa2,strominger} and these singular backgrounds are a limit of regular black holes. The fact that  both the Euclidean path integral approach\footnote{Note that in string theory the presence of the tachyon-like winding modes of the tachyon wrapped around the Euclidean time which survive GSO projection \cite{tachyon,tachyon2} adds to the known disturbing aspect \cite{waldbook} of the Euclidean approach.} \cite{Gibbons} and the Noether current method \cite{wald,wald2} express the entropy of a given black hole in term of its event horizon is not sufficient to conclude that entropy could not be associated to geometries without the event horizon.  We would like to point out that  Mathur and Lunin's description of the entropy \cite{unknown} may resolve the dilemma in the second way. It is intersting that for the case of singular backgrounds representing D1-D5 branes, which have an entropy associated to them, both  Mathur-Lunin description \cite{Rychkov:2005ji} and the subleading string corrections \cite{Ghodsi:2006cd} can generate the entropy. In this note we study if the inclusion of subleading corrections can generate a horizon for backgrounds representing a fundamental string.\footnote{The subleading string corrections to the Schwarzschild black hole has been studied in \cite{sigma1,Cho:2002hq}.}

\section{The \protect{$\alpha'$} corrections}
String theory provides two kind of perturbative corrections to a given background; the string loop corrections and the string world-sheet ($\alpha'$) corrections. The string coupling constant of (\ref{background1}), $g_s^2=g_0^2 e^{2\phi}$,    is
\begin{eqnarray}
g_s^2&=& g_0^2 \frac{~r^{D-3}}{\sqrt{(r^{D-3}+2W)(r^{D-3}+2N)}}\leq g_0^2\,,
\end{eqnarray}
where $g_0$ is an arbitrary parameter. We choose a sufficiently small value for  $g_0$. Thus we ignore the string loop corrections. The $\alpha'$ corrections to the Lagrangian read
\begin{equation}
L\,=\,L^{\scriptscriptstyle(0)}\,+\,\alpha'\,L^{\scriptscriptstyle(1)}\,+\,\alpha'^2\,L^{\scriptscriptstyle(2)}\,+\,\cdots\,,
\end{equation}
where $L^{\scriptscriptstyle(0)}$ stands for the tree-level Lagrangian and the rest is its successive subleading corrections. This series may not make sense for (\ref{background1}) since each term of the $\alpha'$ series diverges at its singularity. However note that the $\alpha'$ corrections change the background itself
\begin{equation}
g\,\to\,g=g^{\scriptscriptstyle{(0)}}\,+\,\alpha'\,g^{\scriptscriptstyle{(1)}}\,+\,\alpha'^2\,g^{\scriptscriptstyle{(2)}}\,+\,\cdots\,,
\end{equation}
and the $\alpha'$-corrected metric, possibly, can have a horizon outside which the $\alpha'$ expansion makes sense. Also the $\alpha'$ corrections to the string coupling constant may remain finite outside the horizon and the string loop corrections could be ignored consistently. In order to check this possibility we truncate the equations of motion at $O(\alpha'^2)$. Then we study if a exact solution of the truncated equations is a black hole with a regular horizon outside which the higher order $\alpha'$ corrections are perturbative. The perturbative $\alpha'$ corrections can be computed in the following ways
\begin{itemize}
\item From the scattering amplitudes of string on sphere as done in \cite{stringamplitude1,stringamplitude2,scater1}. This method gives the Low Energy Effective action up to a perturbative field redefinition since field redefinitions do not alter the scattering amplitudes.
\item Requiring exact conformal symmetry in the corresponding sigma model as done in \cite{sigma1,sigma2,sigma3,sigma4,sigma4.5,sigma5,sigma4.5}. In this method a regularisation  and a renormalisation scheme should be chosen prior to computing the beta functions. Different schemes are related to each other by a perturbative field redefinition.
\item Calculating the LEE action in the Heterotic  closed string field theory \cite{Okawa:2004ii}. This computation has not yet been accomplished. However it does not fix the perturbative field redefinition ambiguity since there remains the freedom to redefine the fields \cite{Yang:2005rx}.
\end{itemize}
The first two methods give the same action up to a perturbative field redefinition ambiguity as the result of the consistency of the string theory around flat space-time \cite{nice,SigmaEqScater}. The outcome of the last method should be in agreement with those of the former ones. The linear $\alpha'$ corrections in Heterotic theory derived from string amplitude considerations read \cite{stringamplitude1,stringamplitude2}
\begin{eqnarray}\label{LMT}
S_{MT}^{(1)}&=&\frac{1}{32\,\pi}\,\int d^{10}x \sqrt{-\det g}~\frac{\alpha'}{8}\,e^{-2\boldsymbol{\phi}} \boldsymbol{L}_{MT}^{(1)}\,,\\
\boldsymbol{L}_{MT}^{(1)}&=&{\boldsymbol{R}_{klmn}\boldsymbol{R}^{klmn}} 
-\frac{1}{2}\,\boldsymbol{R}_{klmn}\boldsymbol{H}_{p}^{~kl}\boldsymbol{H}^{pmn} +\nonumber\\
&&-\frac{1}{8}\boldsymbol{H}_{k}^{~mn}\boldsymbol{H}_{lmn}\boldsymbol{H}^{kpq}\boldsymbol{H}^{l}_{~pq}
+\frac{1}{24}\boldsymbol{H}_{klm}\boldsymbol{H}^{k}_{~pq}\boldsymbol{H}_{r}^{~lp}\boldsymbol{H}^{rmq}\nonumber\,.
\end{eqnarray}
This action includes all the linear $\alpha'$ corrections for backgrounds composed of the dilaton, the metric and the NS two form. A general field redefinition 
\begin{eqnarray}
\boldsymbol{g}_{ij}&\to& \boldsymbol{g}_{ij}+\alpha'\boldsymbol{T}_{ij}\,,\\
\boldsymbol{B}_{ij}&\to& \boldsymbol{B}_{ij}+\alpha'\boldsymbol{S}_{ij}\,,\\
\boldsymbol{\phi}&\to& \boldsymbol{\phi}-\alpha'\dfrac{\boldsymbol{X}}{2}\,,
\end{eqnarray}
induces a change in $\boldsymbol{L}_{MT}^{(1)}$ of the form \cite{MMcriterion2}
\begin{eqnarray}\label{LMTFIELDREDEFINITION}
\Delta \boldsymbol{L}&=&- \boldsymbol{T}^{ij} (\boldsymbol{R}_{ij}-\frac{1}{4}\boldsymbol{H}_{ikl}\boldsymbol{H}_{j}^{~kl}+2\nabla_i\nabla_j\boldsymbol{\phi})+\\
&&+(\frac{1}{2}\boldsymbol{T}_{i}^{~i}+\boldsymbol{X})(\boldsymbol{R}-\frac{1}{12}\boldsymbol{H}^2+4\nabla^2\boldsymbol{\phi}-4(\nabla\boldsymbol{\phi})^2)-\frac{1}{2}\nabla_{k}\boldsymbol{S}_{lm} \boldsymbol{H}^{klm}\nonumber\,.
\end{eqnarray}
where $X,S_{ij}$ and $T_{ij}$ are tensors with appropriate properties and are polynomials of $\boldsymbol{g}_{ij},\boldsymbol{B}_{ij},\boldsymbol{\phi}$ and their derivatives.\footnote{To compute $\Delta \boldsymbol{L}$ it is enough to remember that $\boldsymbol{g}^{ij}\delta \boldsymbol{R}_{ij}=(\boldsymbol{\nabla}^i\boldsymbol{\nabla}^j-\boldsymbol{g}^{ij}\boldsymbol{\Box})\delta \boldsymbol{g}_{ij}$.\cite{blau}} We consider only a class of the field redefinition ambiguities parameters given by 
\begin{eqnarray}
\boldsymbol{T}_{ij}&=& a \boldsymbol{R}_{ij} + \frac{b}{8} \boldsymbol{H}_{ikl}\boldsymbol{H}_{j}^{~kl}+ (e-12f)\, \boldsymbol{g}_{ij} \boldsymbol{R}+\,f \boldsymbol{g}_{ij} \boldsymbol{H}_{klm}\boldsymbol{H}^{klm}\,,\\
\boldsymbol{X}+\frac{1}{2}\boldsymbol{T}^{i}_{~i}&=& (c-12f) \boldsymbol{R} + (\frac{d}{12}+3f) \boldsymbol{H}_{ijk}\boldsymbol{H}^{ijk}\,,\\
\boldsymbol{S}_{ij}&=&0\,,
\end{eqnarray}
where $a,b,c,d,e$ and $f$ are real numbers.  This class of field redefinition alters the linear $\alpha'$ corrected action by
\begin{eqnarray}\label{chosemschemes}
\frac{1}{\alpha'}\,\Delta\boldsymbol{L}&=&-a \boldsymbol{R}_{ij} \boldsymbol{R}^{ij}+ (c-e) \boldsymbol{R}^2 + (\frac{d}{12}-\frac{c}{12}+\frac{e}{4}) \boldsymbol{R}\, \boldsymbol{H}^2 -\frac{d}{144} (\boldsymbol{H}^2)^2 \\\nonumber
&&+ (\frac{a}{4}-\frac{b}{8}) \boldsymbol{H}_{ij}^2\boldsymbol{R}^{ij} + \frac{b}{32} \boldsymbol{H}_{ij}^2\boldsymbol{H}^{2ij}\,+O(\nabla\boldsymbol{\phi})\,,
\end{eqnarray}
where
\begin{eqnarray}
\boldsymbol{H}^2_{ij}&=&\boldsymbol{H}_{ikl}\boldsymbol{H}_{j}^{~kl}\,,\\
\boldsymbol{H}^2&=&\boldsymbol{H}_{ijk}\boldsymbol{H}^{ijk}\,,
\end{eqnarray}
and the derivatives of the dilaton are not written to save space. In the forthcoming computations we do not need them. We require the generalisation of the Einstein tensor to be covariantly divergence free for a trivial dilaton.  Adding this requirement to the linear $\alpha'$ corrections changes it to the first order Lovelock gravity \cite{love} where $(a,c-e)=(\frac{1}{2},\frac{1}{8})$.\footnote{Lovelock gravity \cite{love} is a generalisation of Einstein-Hilbert action where the generalisation of Einstein tensor $G_{ij}$: (1) is symmetric in its indices, (2) is a function of the metric and its first two derivatives, (3) is covariantly divergence free. The linear $\alpha'$ corrections can be chosen to satisfy all these conditions \cite{Zwiebach}. However the higher order $\alpha'$ corrections include also higher derivatives of the metric and can not be rewritten as higher order \cite{wheeler} Lovelock gravity \cite{zumino}.} Thus we set $(a,c)=(\frac{1}{2},\frac{1}{8}+e)$ for which the linear $\alpha'$ corrected action reads
\begin{eqnarray}\label{linearalpha}
S&=&\frac{1}{32\pi}\int d^{10}x\, {\sqrt{-\det \boldsymbol{g}}}\, e^{-2\phi}\, \boldsymbol{L}\\
\boldsymbol{L}&=&\boldsymbol{R}-\frac{1}{12} \boldsymbol{H}^2 + 4 |\boldsymbol{\nabla} \boldsymbol{\phi}|^2 + \alpha' \boldsymbol{L}^{(1)} + \alpha'O(\boldsymbol{\nabla} \boldsymbol{\phi})+ O(\alpha'^2) \\
\boldsymbol{L}^{(1)}&=&\frac{1}{8} \boldsymbol{L}_{GB} +\frac{1}{192}\boldsymbol{H}_{klm}\boldsymbol{H}^{k}_{~pq}\boldsymbol{H}_{r}^{~lp}\boldsymbol{H}^{rmq}-\frac{1}{16}\boldsymbol{R}_{klmn}\boldsymbol{H}_{p}^{~kl}\boldsymbol{H}^{pmn}+\nonumber\\
&&+(\frac{b}{32}-\frac{1}{64}) \boldsymbol{H}^2_{ij} \boldsymbol{H}^{2ij}+(\frac{d}{12} - \frac{e}{6}-\frac{1}{96}) \boldsymbol{R}\, \boldsymbol{H}^2 - \frac{d}{144} (\boldsymbol{H}^2)^2 + (\frac{1}{8}-\frac{b}{8}) \boldsymbol{H}^2_{ij} \boldsymbol{R}^{ij}\nonumber\,
\end{eqnarray}
where $\boldsymbol{L}_{GB}=\boldsymbol{R}_{ijkl}\boldsymbol{R}^{ijkl}-4 \boldsymbol{R}_{ij} \boldsymbol{R}^{ij}+ \boldsymbol{R}^2$ is the Gauss-Bonnet term. In the work \cite{Zwiebach} and some follows works the $\alpha'$ corrections were required not to  produce new extrema for the bi-linear part of the action describing deviation from flat Minkowski space. This criterion, the no-ghost criterion, is questionable since the new extrema are not perturbative in $\alpha'$. The criterion we used produces the same results and is independent of the perturbative behaviour of the $\alpha'$ series. However both of these criteria fail to identify a unique action. The MM-criterion \cite{MMcriterion2,MMcriterion} which provides a unique action does not produce a horizon for (\ref{background1}).

\section{Modification of the singularity}
We presume that there exists an exact $\alpha'$ background in the large dimensions which in the string frame reads 
\begin{eqnarray}\label{ec} 
ds_{\text{exact}}&=& - f(r) dt^2 + dr^2 + g(r) d\Omega_{D-2}^2 \\
\phi&=& \phi(r)\,,\hspace*{2cm} T\,=\,T(r)\,,\\
A^{(1)}_{t}&=&A^{(1)}_{t}(r)\,,\hspace*{1.5cm} A^{(2)}_{t}\,=\,A^{(2)}_{t}(r)\,,
\end{eqnarray}
 the large $r$ limits of which are (\ref{background1}), (\ref{background2}) and  (\ref{background3}). The number of the modified supersymmetry charges\footnote{In LEEA the supersymmetry is realised as the symmetry of the action  therefore, at least, the on shell SUSY constraints needs modification upon the inclusion of the subleading corrections. }  of  this $\alpha'$ exact background should be the same as the number of SUSY charges of the tree-level background.  It is conjectured \cite{sen2} that this $\alpha'$ exact background has a regular event horizon with isometry group of $AdS_2\times S^{D-2}$ whose fields in the vicinity of its horizon can be approximated by
\begin{eqnarray}
ds^2&=&\mathit{v_1}(-\rho^2 d\tau^2 + \frac{d\rho^2}{\rho^2})\,+\,\mathit{v_2}d\,\Omega_{D-2}^2\,,\label{nearhorizonmetric}\\
e^{-2\phi(\rho)}&=&\mathit{s}\,,\label{nearhorizonS}\\
T(\rho)&=&\mathit{T}\,,\label{nearhorizonT}\\
F^{(1)}_{t\rho}&=&\mathit{e_1}\,,\label{nearhorizonf1}\\
F^{(2)}_{t\rho}&=&\mathit{e_2}\,,\label{nearhorizonf2}
\end{eqnarray}
where $\mathit{v_1},\mathit{v_2},\mathit{s},\mathit{T},\mathit{e_1}$ and $\mathit{e_2}$ are constant real ($\mathit{s}, \mathit{T}$ are positive) numbers to be fixed by the equations of motion and the behaviour of the fields at infinity. A concrete proof or refutal of this conjecture requires knowing  all the $\alpha'$ corrections. Neither the string scattering amplitudes nor the sigma model techniques nor CSFT are practically useful to compute the infinite terms of the $\alpha'$-expansion series. There exists no other known method capable of producing the full $\alpha'$-corrected action.\footnote{There have been attempts to guess a compact form for the $\alpha'$ expansion series of the metric \cite{Wohlfarth,Grumiller}.} Currently the conjecture is supported by
\begin{enumerate}
\item Inclusion of only the Gauss-Bonnet action in the induced action allows for the existence of a local horizon geometry whose modified thermodynamical entropy \cite{wald,wald2,Myers}  is in agreement with the statistical entropy up to a numerical constant \cite{sen2}.
\item Inclusion of $R_{ijkl}R^{ijkl}$ and the terms needed by SUSY \cite{19,20,21,22,23,24,25,26,27,28} in the four dimensional  induced action allows for a local horizon whose modified thermodynamical entropy is in agreement with the statistical entropy \cite{atish2}. In higher dimensions it is not known which terms should be added to $R_{ijkl}R^{ijkl}$ to maintain SUSY. 
\end{enumerate}
The conjecture may be contradicted by :
\begin{enumerate}
\item The fundamental string is a special case of the null sigma models \cite{Horowitz:1994rf,Cvetic:1995yq}. It means that there exists a scheme in which the background fields retain their forms at the supergravity approximation. Thus within this scheme the fundamental string remains as a null singular background even after the inclusion of all the $\alpha'$ corrections. Does this contradict the appearance of a horizon due to the inclusion of the $\alpha'$ corrections? 
\item  The value of Wald's entropy is invariant under field redefinition provided that  the $\alpha'$ terms are studied as perturbations around a classical background \cite{Jacobson:1993vj}. Here since Wald's formula is applied on the local horizon which is the exact solution of the truncated equations of motion then Wald's entropy depends on the field redefinition ambiguity parameters. Therefore which values should be chosen for the field redefinition parameters to calculate Wald entropy?
\item The Gauss-Bonnet action or the supersymmetric version of curvature squared terms are not all the linear $\alpha'$ corrections. This fact was also pointed out in \cite{Sahoo:2006rp}.  Does the inclusion of all the linear $\alpha'$ corrections allow for the existence of the horizon?
\item Is there a smooth interpolating solution from the horizon toward the asymptotic infinity? 
\item Could the higher order $\alpha'$ corrections be consistently neglected?
\end{enumerate}
Let us consider the $\alpha'$ expansion series for the Lagrangian density,
\begin{eqnarray}
\mathcal{L}(p)&=&\sum_{n=0}^{\infty}\,\alpha'^n\,\mathit{L}_n(p) 
\end{eqnarray}
where $p$ represents a point in the space-time on which the Lagrangian density is evaluated  and $L_0(p)$ is the Lagrangian density at the supergravity approximation and $\mathit{L}_n(p)$ is the $n^{\text{th}}$ order $\alpha'$ corrections to the Lagrangian density at the supergravity approximation. There exist neighbourhoods around the asymptotic infinity where $\sum\alpha'^n\,\mathit{L}_n(p)$ is an absolute convergent series. We call the union of all these neighbourhoods as the $\cal{C}$-neighbourhood. We refer to the boundary of the $\cal{C}$-neighbourhood as the $\cal{C}$-horizon. The $\cal{C}$-neighbourhood defines a subset of the space-time in which $\sum\,\alpha'^n\,\mathit{L}_n(p)$ is defined unambiguously in the sense that the rearrangements of terms in $\sum\alpha'^n\,\mathit{L}_n(p)$ does not change the series sum, $\mathcal{L}(p)$. The singularity of the supergravity approximation is outside of the $\cal{C}$-neighbourhood. In general the $\alpha'$ corrections could be positive or negative. This means that there exist neighbourhoods in which $\sum\alpha'^n\,\mathit{L}_n(p)$ is a conditionally convergent series. We refere to the union of all these neighbourhoods as the $\cal{NC}$-neighbourhood. The $\cal{NC}$-neighbourhood has two boundaries, the $\cal{C}$-horizon is one of them and we call the other boundary as the $\cal{NC}$-horizon\footnote{The $\cal{C}$-horizon and $\cal{NC}$-horizon are scheme dependent. The $\cal{C}$-horizon could be pushed toward infinity by a field redefinition but the $\cal{NC}$-horizon might not shrink to a point under any field redefinition. It is tempting either to identify the boundary of the union of the $\cal{NC}$-neighbourhoods of all the schemes as a mathematical description for the ``stretched horizon'' defined in \cite{unknown} applied to the case of a wrapped fundamental string or to choose the schemes in which the $\cal{NC}$-horizon coincides with the $\cal{C}$-horizon and then to identify the boundary of the union of the $\cal{C}$-neighbourhoods of all such schemes as the ``stretched horizon''.}. The Lagrangian density on the singularity should be defined as the extrapolation of $\sum\alpha'^n\,\mathit{L}_n(p)$ from $\cal{NC}$-neighbourhood toward the singularity. In the $\cal{NC}$-neighbourhood by a suitable rearrangement of terms, $\mathcal{L}(p)$ may be made to converge to any desired values or even diverge. In the number theory this statement sometimes is referred to as the Riemann theorem. The field redefinition can be thought as a tool to ``rearrange'' the $\alpha'$ series. Thus the  Lagrangian density before reaching the singularity of the supergravity approximation depends on the rearrangements of the terms or almost equivalently on the field redefinition ambiguities. We do not know which of these rearrangements would be preferred or chosen by the underlying conformal field theory since it is not known what a conformal field theory (and if a unique one) represents a wrapped fundamental string.  Ref. \cite{Horowitz:1994rf,Cvetic:1995yq} shows that there exists a scheme in which the background fields retain there forms at the supergravity approximation. This does not mean that we could not rearrange the $\alpha'$ expansion series in the $\cal{NC}$-neighbourhood and then extrapolate the Lagrangian density toward the singularity in such a way that the singularity is covered by an $\alpha'$ stretched horizon. In addition we learn that to be consistent the $\alpha'$ stretched horizon should be at least outside the $\cal{C}$-neighbourhood. Thus the $\alpha'$ series on the $\alpha'$ stretched  horizon are not absolutely convergent series. 

We do not know all the $\alpha'$ series. Therefore we could not identify the $\cal{C}$-neighbourhood and the $\cal{NC}$-neighbourhood in order to compare them with the stretched horizon. In the following we include all the linear $\alpha'$ corrections in a general scheme. We truncate the $\alpha'$ series at $O(\alpha'^2)$. We will show that the local horizon exists upon the inclusion of all linear $\alpha'$ corrections. We illustrate that in general the modified Hawking-Bekenstein entropy associated to the local horizon is not the same for actions related to each other by field redefinitions. Amongst these actions, the choices for which the modified Hawking-Bekenstein entropy is in agreement with the statistical entropy would be preferred.  We provide convincing arguments that the interpolating solution exists and we show that in some schemes the higher order corrections can be ignored outside the $\alpha'$ stretched horizon.

We obtain the linear $\alpha'$ corrections to the induced action by applying the compactification process to the linear $\alpha'$ corrected action in ten dimensions (\ref{linearalpha}). We consider the linear $\alpha'$ corrected action in (\ref{linearalpha}) for all values of the field redefinition parameters,  $(b,d,e,f)$. Ref. \cite{exir3} shows that the pull back of (\ref{LMTFIELDREDEFINITION}) to the four dimensional space time is a functional of the  gauge field strengths of $A_1$ and $A_2$. Thus we can use the entropy formalism \cite{int5,sen0} to express the local horizon parameters in terms of the electric charges. The entropy formalism defines the entropy function by 
\begin{eqnarray}
f(\vec{\mathit{v}},\mathit{T},\vec{\mathit{e}}) &=& \frac{1}{32\,\pi}\, \int\,d\theta\,d\phi\,\sqrt{-\det g}\,\mathit{s}\,L(\vec{\mathit{v}},\mathit{T},\vec{\mathit{e}}) 
\end{eqnarray}
where $L(\vec{v},T,\vec{e})$ is the induced Lagrangian evaluated on the horizon configuration,
\begin{eqnarray}
S&=&\frac{1}{32\,\pi}\,\int d^4x\, \sqrt{-\det g}\, e^{-2\phi}\,L(\vec{\mathit{v}},\mathit{T},\vec{\mathit{e}}). 
\end{eqnarray}
Then the equations of motions are equivalent to 
\begin{eqnarray}\label{eq1}
\frac{\partial f}{\partial \mathit{v_i}} &=&0\,,\\ 
\frac{\partial f}{\partial \mathit{s}} &=&0\,,\\
\frac{\partial f}{\partial \mathit{T}} &=&0\,,\\
\frac{\partial f}{\partial \mathit{e_1}} &=&\frac{N}{2}\,,\\\label{eq5}
\frac{\partial f}{\partial \mathit{e_2}} &=&\frac{W}{2}\,,
\end{eqnarray}
where we have used the notation of Appendix A of \cite{sen0} for the normalisation of the charges.
To evaluate the induced action near the horizon  we employ (\ref{compactification}) to reconstruct the horizon configuration in ten dimensions from (\ref{nearhorizonmetric})-(\ref{nearhorizonf2})\footnote{The compactification of the Gauss-Bonnet action has been done in \cite{ReducedGB}.}
\begin{eqnarray}\label{background10d}
d{\boldsymbol{s}}^2&=&ds^2+\mathit{T}^2 ( dy + 2\,\mathit{e_1}\, r\, d\tau)^2+\sum dz_i^2,\nonumber\\
e^{-2{\boldsymbol{\phi}}} &=& \frac{\mathit{s}}{\mathit{T}}\,,\\
{\boldsymbol{B}} &=&-2\, \mathit{e_2}\, r\, d\tau\wedge\, dy\,.\nonumber
\end{eqnarray}
where the gauges are fixed by
\begin{eqnarray}
A_1 &=& (e_1\, r, 0,0,0)\,,\\ 
A_2 &=& (e_2\, r, 0,0,0)\,.
\end{eqnarray}
Note that the class of  field redefinitions considered in (\ref{chosemschemes}) includes any  field redefinition which produces non-zero terms in the action near the horizon (\ref{background10d}) and whose metric and NS two-form equations of motion are second order differential equations. For the sake of simplicity from this time on we set $D=4$ and we study the four dimensional background representing the fundamental string,
\begin{eqnarray}
D&=&4\,. 
\end{eqnarray}
Using the ten dimensional background near the horizon (\ref{background10d}) one finds that 
\begin{eqnarray}
\mathit{L_0}&=&\boldsymbol{R}-\frac{1}{12}\boldsymbol{H^2}\,=\,-\,\frac{2}{\mathit{v_1}}\,+\,\frac{2}{\mathit{v_1}}
\,+\,\frac{2\,\mathit{e_1}^2\,\mathit{T}^2}{\mathit{v_1}^2}
\,+\,\frac{2\,\mathit{e_1}^2}{\mathit{v_1}^2\,\mathit{T}^2}\\
\mathit{L_1}&=&\frac{1}{8}\boldsymbol{L}_{GB}\,=\,-\,\frac{1}{\mathit{v_1}\,\mathit{v_2}}
\,+\,\frac{T^2\,\mathit{e_1}^2}{\mathit{v_1}^2\,\mathit{v_2}}\,\\ 
\mathit{L_2}&=&\frac{1}{192}\boldsymbol{H}_{klm}\boldsymbol{H}^{k}_{~pq}\boldsymbol{H}_{r}^{~lp}\boldsymbol{H}^{rmq}\,=\,
\frac{\mathit{e_2}^4}{2\,\mathit{v_1}^4\,\mathit{T}^4}\\
\mathit{L_3}&=&-\frac{1}{16}\boldsymbol{R}_{klmn}\boldsymbol{H}_{p}^{~kl}\boldsymbol{H}^{pmn}\,=\,
\frac{\mathit{e_1}^2\,\mathit{e_2}^2}{\mathit{v_1}^4}\,-\,\frac{\mathit{e_2}^2}{\mathit{v_1}^3\,\mathit{T^2}}\\ 
\mathit{L_4}&=&(\frac{b}{32}-\frac{1}{64})\,\boldsymbol{H}^2_{ij} \boldsymbol{H}^{2ij}\,=\,6\,
(b-\frac{1}{2})\frac{\mathit{e_2}^4}{\mathit{v_1}^4\,\mathit{T}^4}\\ 
\mathit{L_5}&=&(\frac{1}{8}-\frac{b}{8}) \boldsymbol{H}^2_{ij} \boldsymbol{R}^{ij}\,=\,
2\,(b-1)\,(\frac{\mathit{e_1}^2\,\mathit{e_2}^2}{\mathit{v_1}^4}\,-\,\frac{\mathit{e_2}^2}{\mathit{v_1}^3\,\mathit{T}^2})\\
\mathit{L_6}&=&(\frac{d}{12}-\frac{e}{6}-\frac{1}{96})\boldsymbol{R}\, \boldsymbol{H}^2\,=\,
h\,\mathit{e_2}^2\,(\frac{1}{\mathit{v_1^3}\,\mathit{T}^2}\,-\,\frac{\mathit{e_1}^2}{\mathit{v_1}^4}-\frac{1}{\mathit{v_1}^2\,\mathit{v_2}\,\mathit{T}^2})\\ 
\mathit{L_7}&=&\frac{d}{144}\,(\boldsymbol{H}^2)^2\,=\,4\,d\,\frac{\mathit{e_2}^4}{v_1^4\,T^4}\,
\end{eqnarray}
where we used $h$ defined by $h\,=\, 4 d - 8 e - \frac{1}{2}$ to represent $\mathit{L_6}$ in a more convenient way. Inserting the above expressions in ten dimensional action we get
\begin{eqnarray}
S&=&\boldsymbol{S}\,=\,\frac{1}{32\, \pi}\, \int \,dt\,dr\,d\phi\,d\cos\theta\,\mathit{s}\,\mathit{v_1}\,\mathit{v_2}\,(\mathit{L_0}\,+\,\alpha'\,\sum_{i=1}^{7}\mathit{L_i})+O(\alpha'^2)\,, 
\end{eqnarray}
where the integration over the compact space is understood. Then the entropy function follows
\begin{eqnarray}\label{insteadofusingwhich}
f(\vec{\mathit{v}},\vec{\mathit{e}},\mathit{s},\mathit{T})&=&\frac{1}{8}\,\mathit{s}\,\mathit{v_1}\,\mathit{v_2}(\mathit{L_0}\,+\,\alpha'\,\sum_{i=1}^{7}\mathit{L_i})  
\end{eqnarray}
where we have truncated the $\alpha'$ series. Using (\ref{insteadofusingwhich}) in (\ref{eq1})-(\ref{eq5}) gives the equations of motion. The solution of the equations of motion identifies the horizon parameters. The identification of the near horizon geometry of half BPS backgrounds  is an example of the supersymmetric attractor mechanism \cite{Ferrara:1995ih,Strominger:1996kf}, where the explicit equations of motion are solved rather than the supersymmetric constraints. Solving the equations of motion was first carried out by Ashoke Sen in \cite{sen2} where only the Gauss-Bonnet Lagrangian was included in the induced action. The Gauss-Bonnet Lagrangian in the four dimensions reads
\begin{eqnarray}
\frac{1}{8}(R_{ijkl}R^{ijkl}-4 R_{ij} R^{ij} + R^2) &=& -\frac{1}{\mathit{v_1}\,\mathit{v_2}} 
\end{eqnarray}
which coincided with the first term in $\mathit{L_1}$. We see that in total five terms in the the summation of $L_1+\cdots+L_7$ are not reproduced by the inclusion of the  four-dimensional Gauss-Bonnet Lagrangian.

A linear combination of the  equations of motion of $T$  and  of $v_1$  factorises
\begin{eqnarray}
\dfrac{\partial f}{\partial \mathit{s}}&=&0 ~\to f\,=\,0\,,\\ \label{factorf1}
\left. (\frac{1}{\mathit{T}}\dfrac{\partial f}{\partial \mathit{T}}-4 \mathit{e_1}^2 \dfrac{\partial f}{\partial \mathit{v_1}})\right.|_{_{{f}=0}} &=& (\mathit{T}^2 \mathit{e_1}^2 - \frac{\mathit{v_1}^2}{4}) (...)\,.
\end{eqnarray}
Eq. (\ref{factorf1}) implies that some of the solutions may be given by 
\begin{eqnarray}\label{f1found}
e_1&=& \dfrac{\sqrt{v_1}}{2T} \,.
\end{eqnarray}
 Eq. (\ref{f1found}) simplifies the equations of motion of $\mathit{v_1},\mathit{v_2},\mathit{s}$ and $\mathit{T}$ and enables one to solve them,
\begin{eqnarray}
\mathit{v_1}&=& (3+ h\, x^2)\,\frac{\alpha'}{8}\,,\label{v1found}\\
\frac{\mathit{v_2}}{\mathit{v_1}}&=& \frac{4(1+ h\, x^2)}{-\,h\,x^4 + (3h+4b-5) x^2 + 15}\,,\label{v2found}\\\label{Sfound}
\mathit{s} &=&\sqrt{\frac{x\,N\,W}{v_1}}\,\frac{h\,x^4\,+\,1}{3\,+\,(b-1)\,x^2}\,\frac{\mathit{v_1}}{\mathit{v_2}}\\
\mathit{T}&=&\sqrt{\frac{N}{W\,x}}\label{Tfound}\\
\mathit{e_2}&=& \frac{1}{2}\sqrt{v_1}\, x\, T\,,\label{f2found}
\end{eqnarray}
where $x$ is a root of
\begin{eqnarray}\label{xfound}
 (-4\,d\,-\,6\,b\,-\,h\,+\,\frac{5}{2})\, x^4\, -\, 6\,(1\,-\,b)\,x^2\,+\, 9 &=&0\,,
\end{eqnarray}
Note that  we used $x$  as a different parametrisation of $b,d,h$ to express the near horizon configuration in a more convenient way. Eq's (\ref{f1found})-(\ref{f2found}) identify the near horizon configuration. We use the entropy formula of entropy formalism \cite{int5,sen,sen0} to calculate Wald's entropy associated to the local horizon. The entropy formalism expresses Wald's entropy, $S_{BH}$, by
\begin{eqnarray}
S_{BH}&=& 2 {\pi} (\frac{\partial{f}}{\partial \mathit{e_1}} \mathit{e_1}+\frac{\partial{ f}}{\partial \mathit{e_2}} \mathit{e_2}- {f}), 
\end{eqnarray}
which is evaluated on the horizon. We can use (\ref{eq1})-(\ref{eq5}) to write
\begin{eqnarray}\label{simpleSBH}
S_{BH}&=&  2\,{\pi} (\frac{N}{2}\, \mathit{e_1}+\frac{W}{2}\, \mathit{e_2})\,=\,\pi\,\sqrt{N\,W\,x\, \mathit{v_1}} 
\end{eqnarray}
where we used the local horizon parameters (\ref{f1found}),  (\ref{Tfound}) and  (\ref{f2found}). We see that both the local horizon parameters and the entropy depend on the field redefinition ambiguity parameters. We have expected this dependence since we have applied Wald's entropy formula on the exact solution of the truncated action. The equality of the statistical entropy (\ref{statisticalentropy}) and Wald entropy (\ref{simpleSBH}) happens in the schemes where
\begin{eqnarray}\label{WaldStatistical}
x\,\mathit{v_1} &=& {\alpha'} 
\end{eqnarray}
and we choose these schemes. There exist a set of ranges for the parameters of the field redefinition ambiguity where $\mathit{v_1},\mathit{v_2},\mathit{T},\mathit{s} $ are all positive. It is straightforward to identify these ranges. Here we focus on the subset of the parameters where identity is a root of (\ref{xfound}) or equivalently $h=-4d+\frac{11}{2}$. In this subset T-duality in the $y$ direction (\ref{ycordinate}) remains trivial in the sense that interchanging $N$ and $W$ describes T-duality both at asymptotic infinity and near the horizon.\footnote{In general requiring T-duality to commute with $\alpha'$ corrections identifies  corrections to T-duality\cite{T1,T2}. The explicit form of the $\alpha'$ corrections to T-duality on backgrounds composed of a diagonal metric and the dilaton is presented in \cite{martin,exir2}.} Then using (\ref{WaldStatistical}) for $x=1$ fixes $d$ to $d=\frac{1}{8}$ for which the near horizon configuration is simplified to
\begin{eqnarray}
\mathit{v_1}&=& 16,\\
\frac{\mathit{v_2}}{\mathit{v_1}}&=& \frac{6}{5}\,,\\
\mathit{T}&=& \sqrt{\frac{N}{W}}\,,\\
\mathit{e_1}&=&2\sqrt{\frac{W}{N}}\,,\\
\mathit{e_2}&=&2\sqrt{\frac{N}{W}}\,,\\
\mathit{s}&=&\frac{5}{8}\,\sqrt{{N\,W}}\,,
\end{eqnarray}
and we have chosen $b=0$ and used the unit of $\alpha'=16$. We see that $(\frac{v_1}{\alpha'},\frac{v_2}{\alpha'})\sim(1,1)$, and the stretched horizon is not larger than $\alpha'$. We can choose other values for the field redefinition ambiguity parameters to make the local horizon arbitrarily large. For example we can choose $x=\frac{1}{2},b=0, h=52, d=\frac{141}{8}$ to get
\begin{eqnarray}
\mathit{v_1}&=& 2\,\alpha',\\
\mathit{v_2}&=& \frac{224}{99}\,\alpha'\,,\\
\mathit{T}&=&\sqrt{2\,\frac{N}{W}}\,,\\
\mathit{e_1}&=&\frac{{1}}{2}\sqrt{\frac{\alpha'\,W}{N}}\,,\\
\mathit{e_2}&=&\frac{{1}}{2}\sqrt{\frac{\alpha'\,N}{W}}\,,\\
\mathit{s}&=& \frac{9}{4}\,\sqrt{\frac{N\,W}{\alpha'}}\,,
\end{eqnarray}
for which one can argue that the higher order $\alpha'$ corrections are suppressed outside the horizon and the higher order $\alpha'$ corrections only provide perturbations around the ``black hole''. We expect that there exist schemes\footnote{\cite{Prester:2005qs} has found the local horizon configuration parameters in a general dimension for a general Lovelock gravity.} in which Wald's entropy for a black hole of a general dimension is in agreement with the statistical entropy and $(\frac{v_1}{\alpha'},\frac{v_2}{\alpha'})>(1,1)$, therefore, the higher order $\alpha'$ corrections could be ignored outside the stretched horizon within these schemes. However we notice that the values of the field redefinition parameters are not small in these schemes. For the case of the WZW models where the exact conformal theory is known the values of the field redefinition ambiguity in which the background fields retain their forms are at order one \cite{Sfetsos:1993ka}. Thus it is unlikely that the large values for the field redefinition ambiguity parameters are going to be chosen by the underlying conformal field theory.

Note that there exist field redefinition ambiguities which vanish near the horizon and infinity. The class of field redefinitions that leave the  equations of the metric and NS two-form as second order differential equations is 
\begin{eqnarray}
 T_{ij}&=& c_1 \, \boldsymbol{\nabla}_i\boldsymbol{\nabla}_j \boldsymbol{\phi} + c_2 \boldsymbol{g}_{ij }\boldsymbol{\Box} \boldsymbol{\phi} + c_3 \boldsymbol{\nabla}_i\boldsymbol{\phi}\boldsymbol{\nabla}_j\boldsymbol{\phi}+c_4 \boldsymbol{g}_{ij}|\boldsymbol{\nabla}\boldsymbol{\phi}|^2 \\
X &=& c_5 \boldsymbol{\Box} \boldsymbol{\phi} + c_6 |\boldsymbol{\nabla} \boldsymbol{\phi}|^2
\end{eqnarray}
where $c_1, c_2, \cdots,c_6$ are arbitrary real numbers. Ref. \cite{sen2,Hubeny:2004ji} have looked for a numerical interpolating solution in one single set of the ambiguity parameters. One should study if there exists any set of values for $b,d,e,f,c_1,\cdots,c_6$ for which a smooth solution interpolates from the near horizon geometry to infinity. This question needs further investigation, however  due to the large numbers of the free parameters it is tempting to argue that the interpolating solution exists in general.

\section{Conclusions}
We have studied the linear $\alpha'$ corrections and the field redefinition ambiguities in the critical Heterotic string theory for the backgrounds representing a  fundamental string wrapped around a two cycle.

We have required the $\alpha'$ corrections to the Einstein tensor to be covariantly divergence free. This requirement has enabled us to rewrite the square of the Riemann tensor as the Gauss-Bonnet Lagrangian keeping some of the field redefinition ambiguity parameters untouched. One may ask if this requirement, similar to the ghost-freedom criterion \cite{GhostFreedom}, could be applied to all orders in $\alpha'$. This question needs further investigation. It would be intersting to find a criterion which both fixes the remaining ambiguity parameters and gives rise to a stretched horizon for half-BPS singular backgrounds representing a wrapped fundamental string. The MM-criterion \cite{MMcriterion2,MMcriterion} which provides a unique action does not produce the stretched horizon.

We have included all the linear $\alpha'$ corrections for half-BPS singular backgrounds representing a wrapped fundamental string. We have studied all the schemes, field redefinitions ambiguities, compatible with the ten dimensional diffeomorphism group. We have shown that there exist schemes in which the inclusion of all the linear $\alpha'$ corrections gives rise to a `local' horizon with geometry $AdS_2\times S^{D-2}$ and for which the modified Hawking-Bekenstein entropy is in agreement with the statistical entropy. Note that the modified Hawking-Bekenstein entropy is the same for actions related to each other by field redefinition provided that the $\alpha'$ terms are studied as perturbations around a classical solution \cite{Jacobson:1993vj}. However since we have applied Wald's formula on the exact solution of the truncated equations of motion, the entropy depends on the field redefinition ambiguity parameters.

We have shown that there exist schemes in which the $\alpha'$ stretched horizon is large  and Wald's entropy is in agreement with the statistical entropy. Thus  within these schemes the higher order $\alpha'$ corrections can be ignored outside the stretched horizon. Also we have argued that a smooth solution connects the $\alpha'$ stretched horizon to the fall off of the fields at asymptotic infinity.

This means that there exist schemes in which the $\alpha'$ stretched horizon is small and also there exist schemes where the $\alpha'$ stretched horizon does not exist at all. We do not know which scheme would be preferred or chosen by the underlying conformal field theory since it is not known what a conformal field theory (and if a unique one) represents a wrapped fundamental string. Ref. \cite{Horowitz:1994rf,Cvetic:1995yq} shows that there exists a scheme in which the fields of the fundamental string background retain their forms at the supergravity approximation, thus within this scheme the background remains as a null singular background under the inclusion of all $\alpha'$ corrections. We have concluded from this that the $\alpha'$ expansion series is not an absolutely convergent series on the $\alpha'$ stretched horizon whenever the scheme admits the $\alpha'$ stretched horizon.

Although we have proved the existence of the schemes in which the $\alpha'$ stretched horizon is larger than the string length and for which the statistical entropy is in agreement with Wald entropy, still we find disturbing that the the thermodynamical entropy is scheme-dependent. This dependence beside  not absolutely convergent characteristic of the $\alpha'$ series on the $\alpha'$ stretched horizon  may be counted on as indications to express the thermodynamical properties in term of something else, in place of the event horizon, which null-singular geometries possess instead of requiring the subleading corrections to covert the null singular backgrounds to black holes with a regular event horizon. We would like to point out that Mathur and Lunin description for the entropy \cite{unknown} may be employed to generate a thermodynamical entropy for a wrapped fundamental string without first requiring the $\alpha'$ corrections to produce an event horizon covering the singularity.

 \section*{Acknowledgements}
I would like to thank Loriano Bonora and Martin O'Loughlin for  useful discussions and suggestions throughout the work, and careful reading of the draft. I thank  Ashoke Sen and Marco Serone for their comments when I was presenting an early version of this work and  Predrag Prester for correspondence. I thank Justin David for discussions.

\providecommand{\href}[2]{#2}\begingroup\raggedright

\endgroup

\end{document}